\documentclass[a4paper,twoside,10pt]{letter}

\usepackage{graphicx,saj,multicol,subeqnarray}
\usepackage{csvsimple}
\usepackage{pgfplotstable}
\usepackage{lscape}
\usepackage{longtable}
\usepackage{ gensymb }

\usepackage{ wasysym }
\usepackage{ gensymb }
\usepackage{rotating}
\usepackage{pdflscape}

\newcommand{\M}{M\,31}

\def\p0{\phantom{0}}
\def\degr{\hbox{$^\circ$}}
\def\arcmin{\hbox{$^\prime$}}
\def\arcsec{\hbox{$^{\prime\prime}$}}
\newcommand{\farcs}{\mbox{\ensuremath{.\!\!^{\prime\prime}}}}

\def\papertype{Editorial}

\def\udc{...}
\begin{document}
\baselineskip=3.1truemm
\columnsep=.5truecm
\newenvironment{lefteqnarray}{\arraycolsep=0pt\begin{eqnarray}}
{\end{eqnarray}\protect\aftergroup\ignorespaces}
\newenvironment{lefteqnarray*}{\arraycolsep=0pt\begin{eqnarray*}}
{\end{eqnarray*}\protect\aftergroup\ignorespaces}
\newenvironment{leftsubeqnarray}{\arraycolsep=0pt\begin{subeqnarray}}
{\end{subeqnarray}\protect\aftergroup\ignorespaces}
%


\markboth{\eightrm 20~cm VLA Radio-Continuum Study of {\M\ }} {\eightrm T. J. Galvin, M.~D. Filipovi\'c}

{\ }

\publ

\type

{\ }


\title{20~cm VLA Radio-Continuum Study of {\M\ }-- Images and Point Source Catalogues DR2: Extraction of a supernova remnant sample}


\authors{T. J. Galvin, M.~D. Filipovi\'c}

\vskip3mm


\address{$^1$University of Western Sydney, Locked Bag 1797, Penrith South DC, NSW 2751, Australia}

\Email{136525304}{student.uws.edu.au, m.filipovic@uws.edu.au}


\dates{May 5, 2014}{September 2, 2014}


\summary{We present Data Release~2 of the Point Source Catalogue created from a series of previously constructed radio-continuum images of \M\ at $\lambda$=20~cm ($\nu$=1.4~GHz) from archived VLA observations. In total, we identify a collection of 916 unique discrete radio sources across the field of \M. Comparing these detected sources to those listed by Gelfand et al. (2004) at $\lambda$=92~cm, the spectral index of 98 sources has been derived. The majority (73\%) of these sources exhibit a spectral index of $\alpha <$--0.6, indicating that their emission is predominantly non-thermal in nature, which is typical for background objects and Supernova Remnants (SNRs). Additionally, we investigate the presence of radio counterparts for some 156 SNRs and SNR candidates, finding a total of only 13 of these object in our images within a 5\arcsec\ search area. Auxiliary optical, radio and X-ray catalogs were cross referenced highlighting a small population of SNR and SNR candidates common to multi frequency domains.  }



\begin{multicols}{2}
{


\section{1. INTRODUCTION}

As a member of the Andromeda constellation, \M\ is the closest spiral galaxy to our own at a distance of $\sim778$~Kpc (Karachentsev et al.~2004). For this reason, it plays a significant role in galactic and extragalactic studies. A number of previous radio-continuum studies at $\lambda$=20~cm (Braun 1990a) focused on general properties of \M, such as its structure and magnetic fields. Also, Braun (1990b) presented a list of 20~cm (5340 sources in the north-east parts of \M. A number of other studies such as Dickel et al. (1982) estimated flux densities of \M\ supernova remnants (SNRs) and H\textsc{ii} regions. Lee et al. (2014) identified 76 new SNRs based on H$\alpha$ and [S\textsc{ii}] images of \M, and further confirmed a total of 80 SNR candidates from previous literature based on their selection criteria.  

In this paper, we release our first revision of the data catalogue first published in Galvin et al. (2012). The original catalogue was produced from archive Very Large Array (VLA) data that was accessed and imaged from the National Radio Astronomy Observatory (NRAO) online data retrieval system at $\lambda$=20~cm. Here, we also investigate the SNR component of our sample of radio sources in the \M.

\section{2. DATA AND IMAGE CREATION}

A collection of existing, archived radio-continuum observations at $\lambda$=20 cm with pointings centred on \M\ were obtained from the National Radio Astronomy Observatory (NRAO)\footnote{https://archive.nrao.edu/archive/e2earchivex.jsp} online data retrieval system. In total, 15 VLA projects with a variety of array configurations were selected for use in this study, as summarised in Table~1 of Galvin et al. (2012). These projects were observed between the 1$^{st}$ of October 1983 and 27$^{th}$ of September 1997 and are comprised of 28 individual observational runs. 

The \textsc{miriad} (Sault et al. 1995) and \textsc{karma} (Gooch 1996) software packages were used for data reduction and analysis. Initially, observations were loaded into \textsc{aips} and had their source coordinates converted from the B1950 to the J2000 reference frame. They were then exported to \textsc{fits} files so that they could be loaded into the \textsc{miriad} software package, which was then used to perform actual data reduction. Typical calibration, flagging and imaging procedures were then carried out (Sault et al. 1995). For more information on data analysis and image creation, as well as the complete set of final images used for source identification, see Galvin et al. (2012), Galvin et al. (2014), O'Brien et al. (2013) and Payne et al. (2004a).

\section{3. RESULTS}

Using the images produced in Galvin et al. (2012), we re-evaluate the unique source catalogue produced through that work. Table~1 lists all sources and the images in which they were found. Using a 2\arcsec\ search radius, this source catalogue from Galvin et al. (2012) was internally cross referenced to identify sources which were found in more then one image. In such cases, the collection of sources was given an unique group identification number. For each of these groups, an average flux and associated error was calculated based on the individual measurements of each source in that particular group. In total, 916 unique sources were identified, and of which 109 were found in at least two images. 

In Fig.~1 we present the distribution of flux densities of all unique point sources. For unique sources found in more then one image (a group), that groups average flux was used. Of the total 916 unique sources, 882 ($\sim$96\%) are with flux density that is below 50~mJy.  From Fig.~1 we estimate that our completeness level is $\sim$2~mJy. 

In Fig.~2 we show the distribution of the average flux errors of each group (109 in total). The majority of source groups (70 out of 109) have errors which are less then 30\%. Of the 109 source groups, 17 have an error that is greater then 50\%, possibly indicating a population of transient sources in the field of \M. 

Our catalogue was then compared to Gelfand et al. (2004). They identified 405 sources at $\lambda$~=~92cm though a VLA survey conducted in an A type configuration. Their catalogue also focused predominately on compact radio sources, as the longer baselines of the A type configuration resolved out most of the extended emission. Using a 5\arcsec\ tolerance, a total of 98 sources were found to be common between both catalogues. We estimate a spectral index ($ S_\nu \propto \nu^{\alpha}$), for these 98 sources using flux density measurements at $\lambda$=20 and 92~cm. In the event where a $\lambda$=92~cm source was matched to a group of sources from this study (ie. a source found in more then one image), that groups average flux was used in the derivation of that sources spectral index. The distribution of these spectral indices is presented in Fig.~3. These sources exhibit a predominately negative spectral index, indicating a significant population of background sources and/or SNRs in \M\ field.

\section{4. SNRs IN THE \M}

Magnier et al. (1995) presented a comprehensive study of optically identified SNR and SNR candidates in the field of \M. Despite relatively poor seeing conditions ($>2''$), short exposure times and imaging only a portion of \M's complete field, a total of 179 SNR and SNR candidates were identified. An additional 55 SNR and SNR candidates were also included in Magnier et al. (1995) from earlier work by Dodorico et al. (1980), Blair et al. (1981) and Braun and Walterbos (1993). These 234 SNR and SNR candidates were reviewed individually by Lee et al. (2014). Of the 239 SNR candidates presented by Magnier et al. (1995), 154 were discounted by Lee et al. (2014) using their SNR selection criteria.

We compared our radio-continuum catalogue to Lee et al. (2014), who used H$\alpha$ and [S\textsc{ii}] images of \M\ from the Local Group Survey (LGS; Massey et al. 2006) to identify SNR candidates. Using a search criteria of S[\textsc{ii}]/H$\alpha > $ 0.4, morphology and absence of blue stars, they identified a total of 156 SNRs and SNR candidates. Of these 156, some 80 had been listed in previous studies. We searched for these 156 objects in our own catalogue using a 5\arcsec\ search radius. A total of 13 optical SNR candidates from Lee et al. (2014) were found to have radio counter-parts. Some 3 of these 13 were also detected in Gelfand et al. (2004) listed sources. These three sources, J004102+410427, J004339+412653 and J004047+405525, have spectral indices of --0.87, --0.75 and --0.31 respectively. Two SNRs with steeper spectrum are indicative of younger SNR age. Although the relatively steep spectral index of J004102+410427 is more indicative of background galaxy, this source is likely to be intrinsic to \M\ given that it lies on the ring of diffuse emission surrounding the galaxy.
We also compared positions from our radio-continuum catalogue presented here and Lee et al. (2014) and found no significant discrepancy (see Fig.~4). The standard deviations for $\Delta$RA and $\Delta$Dec of these 13 sources in common are 1\farcs4 and 1\farcs3 respectively, and have been marked in Table~2 with a * (also noted in the tables caption). 

In Fig.~5 we present a radio-continuum image of \M\ that was produced in Galvin et al. (2012) overlaid with the positions of 156 SNRs and SNR candidates identified by Lee et al. (2014). The red circle represents an SNR that had been previously identified through earlier studies, while the blue cross represents new SNRs identified by Lee et al. (2014). We see a large portion of SNRs fall on, or nearby, the ring of emission which surrounds the galaxy. The spatial distribution of these SNR candidates and their coincidence with the ring of diffuse emission implies that there is a higher rate of star formation in the outer arms of the galaxy then in its central region. Sources common between our study and Lee et al. (2014) have been marked with a purple star. 

We also compared our radio-continuum catalogue with Dickel et al. (1982) who observed 10 SNRs in the field of \M\ at $\lambda$=20~cm. Of these 10 SNRs, 7 were \ detected in our VLA images. Once cross-referenced with our own catalogue, we find the 4 strongest detected SNRs listed in Dickel et al. (1982) were present in our own images. These four common sources are J004047+405526, J004513+413616, J004135+410657 and J004513+413616 (Group Identification Number of 92) and note that their fluxes are in good agreement. The other three detected SNRs from Dickel et al. (1982) are below our detection limit.

Using existing optical, radio and X-ray catalogues we investigate the SNR population of \M. A radio-continuum catalogue ($\nu$=1465~MHz) comprised of 58 SNR and SNR candidates was presented by Braun and Walterbos (1993). Of these 58 sources, 24 had a signal to noise ratio above 5$\sigma$ and were included for use in this study. Comparing our catalogue to the 24 Braun and Walterbos (1993) SNRs, we find that there are 8 matches within a 5\arcsec\ search radius. Of the remaining unmatched 16 SNRs some 14 have associated extended emission structure in our images. In Fig.~6 we present a radio-continuum image of \M\ that was produced in Galvin et al. (20120) overlaid with the positions of the 24 Braun and Walterbos (1993) sources with a signal to noise ration above 5$\sigma$. The red circles represent sources common to both, our and Braun and Walterbos (1993) study, while the purple crosses represent sources from Braun and Walterbos (1993) that unmatched when compared to our work. 

Additionally, 47 X-ray SNR and SNR candidates were identified by Sasaki et al. (2012) using the {\em XMM-Newton} telescope. Comparing Sasaki et al. (2012) to our catalogue, we find at total of 11 sources in common.

Fig.~7 is a Venn diagram showing the source intersection of these three catalogues using a search radius of 5\arcsec. It is not surprising that there is only \textless 10\% (13 out of 156) coincidences between optical SNRs and SNR candidates and their potential radio counterparts. While in the Magellanic Clouds the ratio between known radio and optical SNRs are $\sim90\%$ (Bozzetto et al. 2014 in prep.; Haberl et al. 2012; Filipovi\'c et al 2008, 2005; Payne et al. 2008, 2007, 2004b) in other nearby galaxies this ratio is in order of 10\%. For example: NGC\,300 ratio is 14\% (Millar et al. 2011, 2012), NGC\,7793 ratio is 23\% (Galvin at al. 2014 in prep.; Pannuti et al. 2011) and NGC\,55 ratio is 10\% (O'Brien et al. 2013). This may indicate that either our present searches for SNRs in the \M\ are not complete and/or biased towards the optical detection implying optical criteria to be bit more sensitive. 

At the same time, we point out that the various density environments in which SNRs are expanding also have an effect on their detections. While radio observations are biased towards the detections of SNRs embedded in higher density environments, optical searches are more sensitive toward SNRs in low density mediums (see Pannuti et al. 2000). Coupling this with the larger distances to external galaxies (and therefore poorer resolution and sensitivity) our searches for external SNRs are at the moment introducing significant selection effects. Therefore, any comparison of SNRs in external galaxies should be taken with great caution.

\section{CONCLUSION}

We present version two of our point source catalogue of \M\ that was created using archived VLA data at $\lambda$=20~cm. In total, 916 unique sources were identified across 15 various VLA projects. Of these 916 sources, 109 of them were found in at least two different images. The spectral index for 98 sources was derived by comparing our catalogue to Gelfand et al. (2004), who lists sources at $\lambda$=92~cm. Also, we compared the optical SNR search results of Lee et al. (2014) to our radio-continuum maps of \M, showing 13 SNRs and SNR candidates in common to two surveys to reside on the ring of diffuse emission surrounding the galaxy. Auxiliary optical, radio and X-ray catalogs were cross referenced highlighting a small population of SNR and SNR candidates common to multi frequency domains. 

}\end{multicols}

\centerline{\includegraphics[angle=-90,width=0.85\columnwidth, keepaspectratio, clip=true, trim=0cm 0cm 0cm 0.5cm]{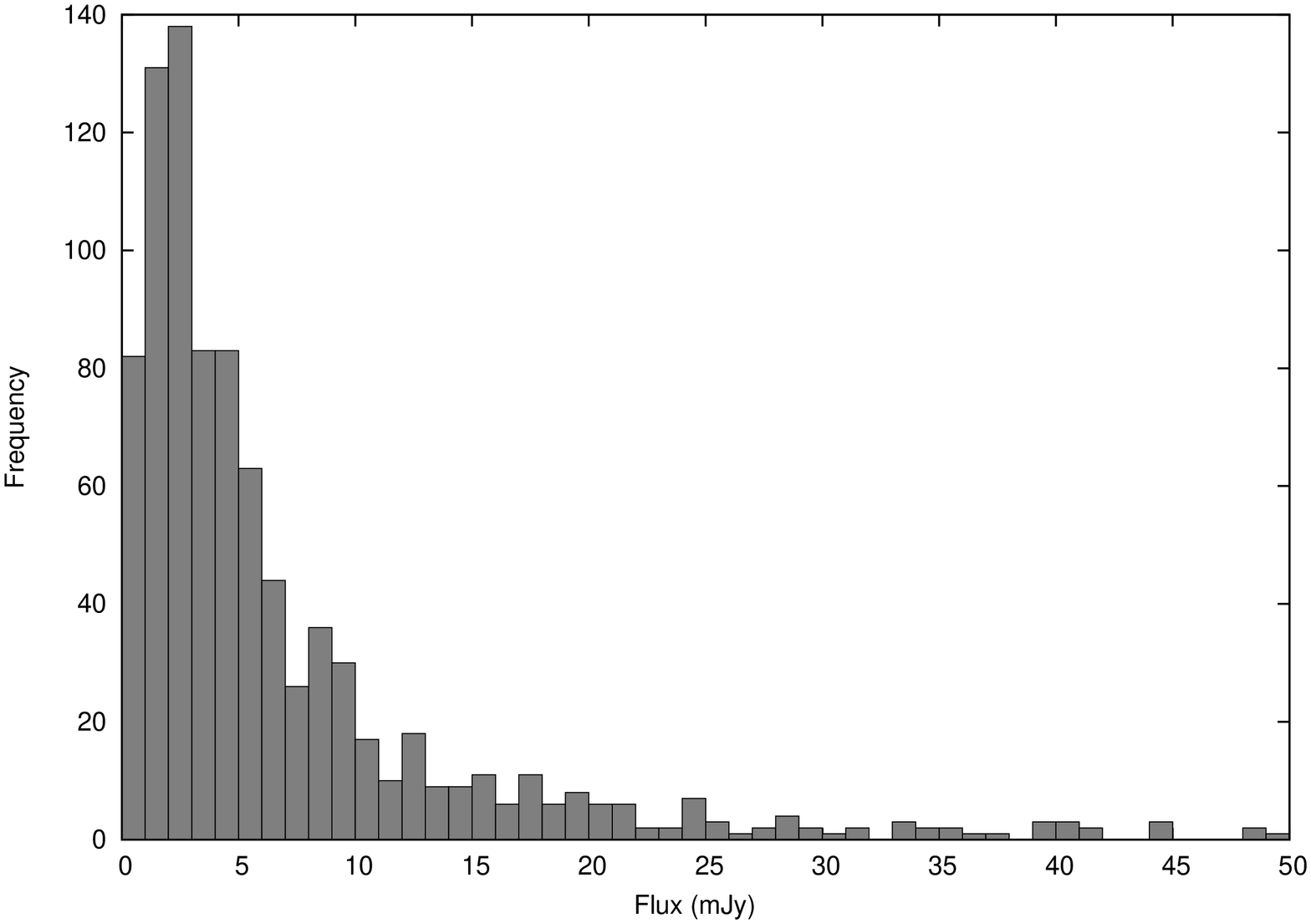}}
\figurecaption{1.}{The distribution of flux densities for 916 radio sources found in \M. 34 sources, whose flux density was above 50~mJy, have been excluded from this graph.}

\centerline{\includegraphics[angle=-90,width=0.85\columnwidth, keepaspectratio, clip=true, trim=0cm 0cm 0cm 0.45cm]{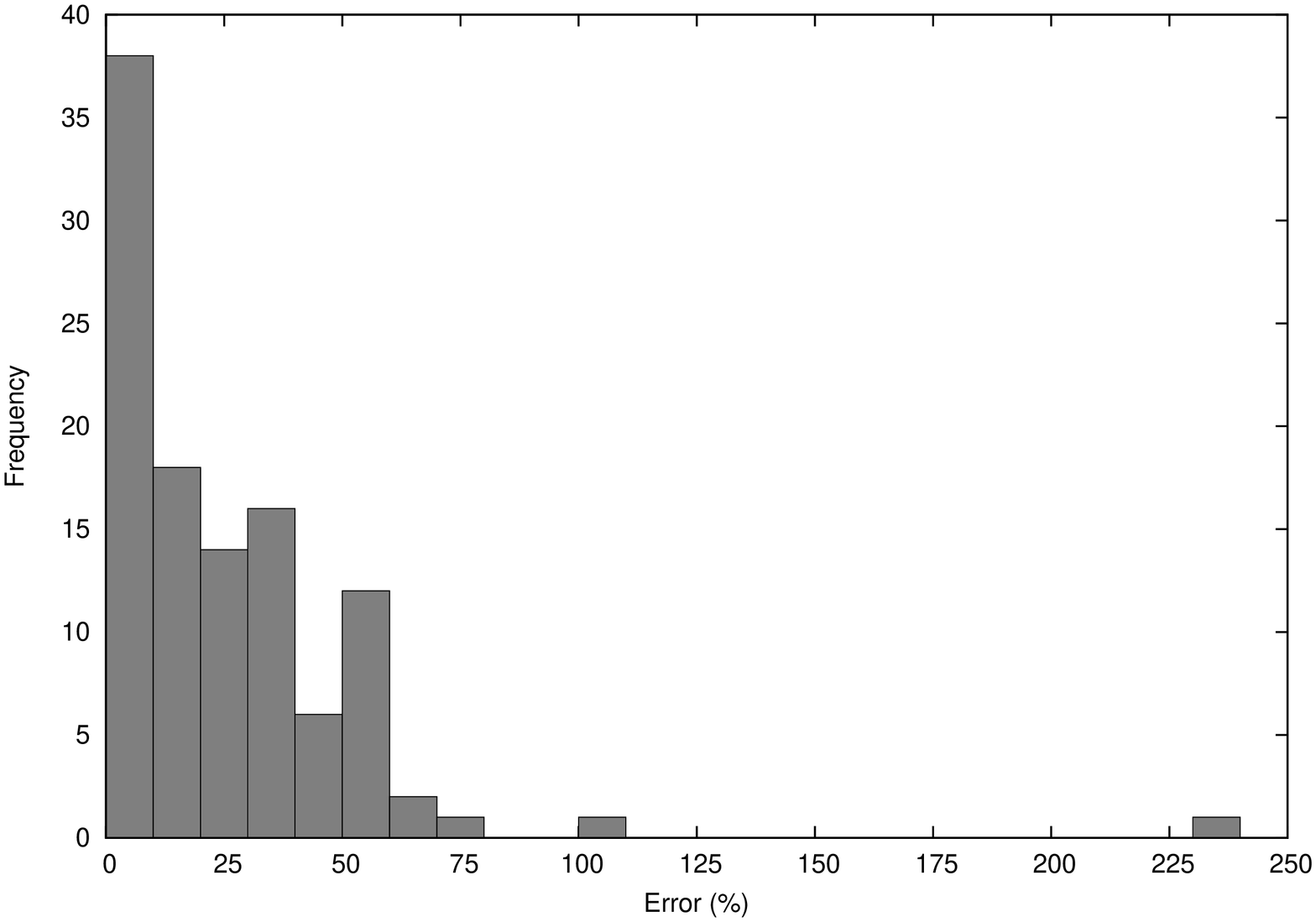}}
\figurecaption{2.}{The distribution of errors for the average flux density in each group of unique sources.}

\centerline{\includegraphics[angle=-90,width=0.85\columnwidth, keepaspectratio, clip=true, trim=0cm 0cm 0cm 0.5cm]{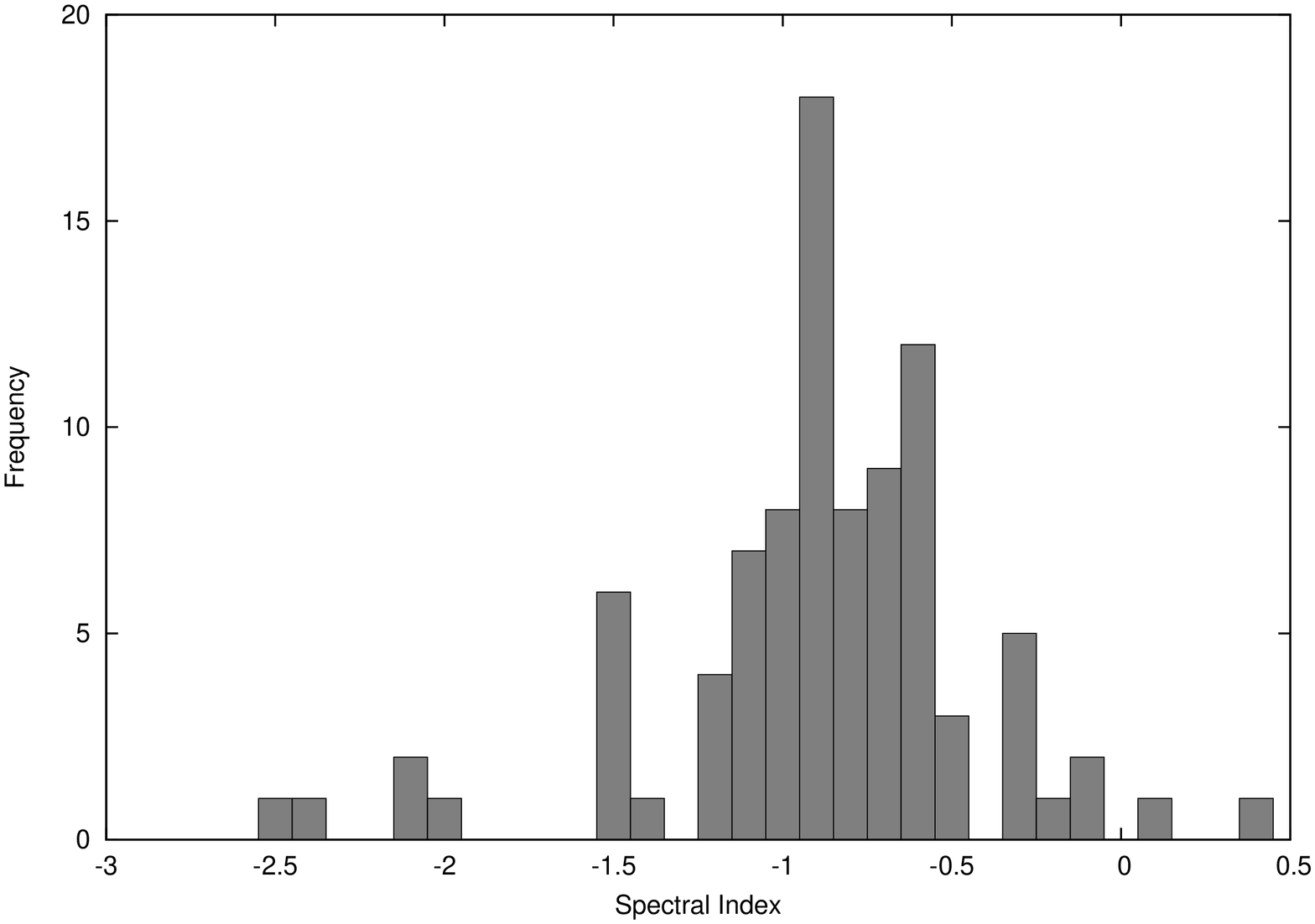}}
\figurecaption{3.}{Spectral index distribution of 98 point sources in the field of \M.\label{fig:spectralhist}}

\centerline{\includegraphics[angle=-90,width=0.8\columnwidth, keepaspectratio]{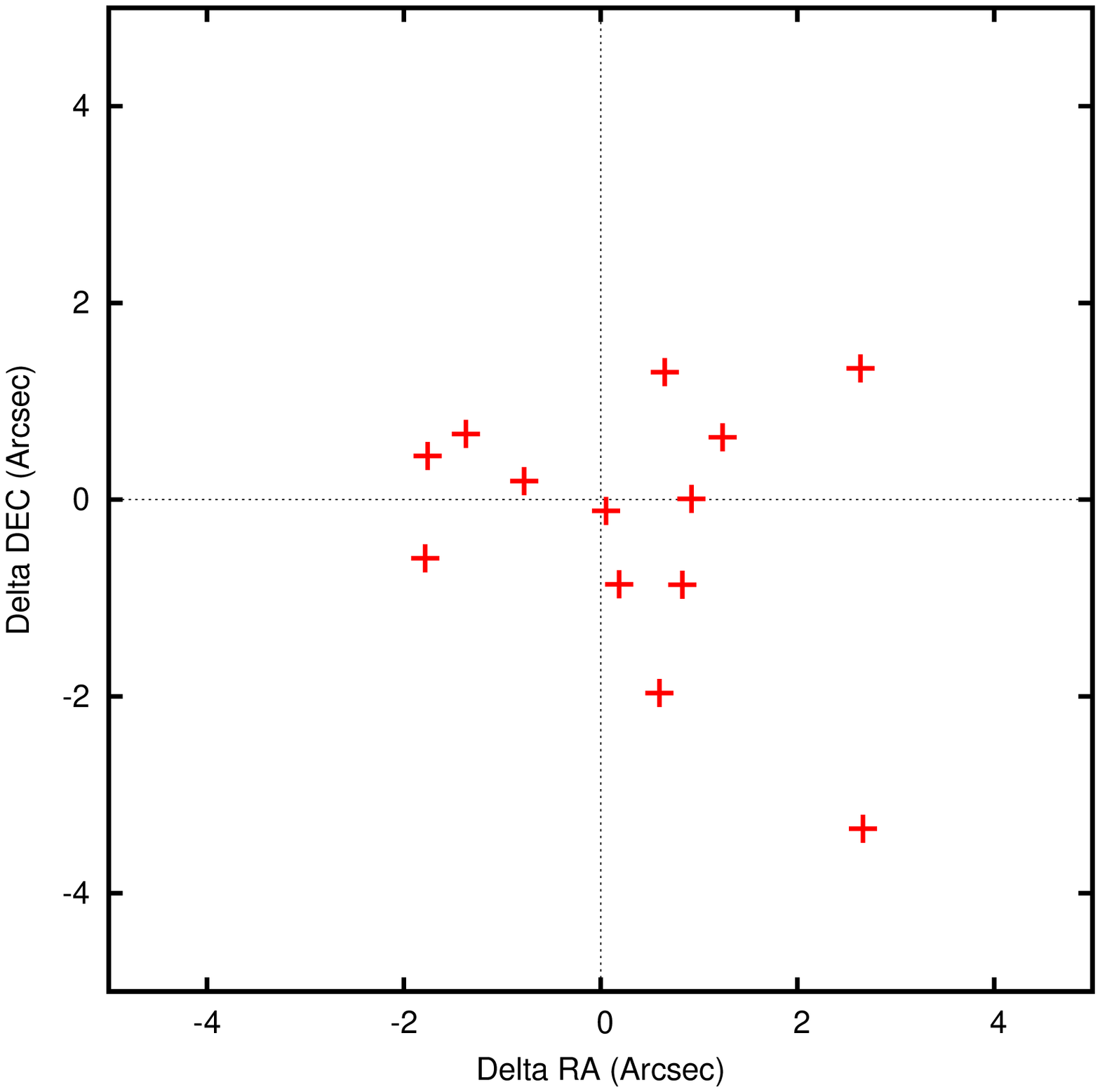}}
\figurecaption{4.}{RA and Dec offsets of 13 SNRs and SNR candidates common between this study and Lee et al. (2014). The standard deviations for $\Delta$RA and $\Delta$Dec are 1.4\arcsec\ and 1.3\arcsec\ respectively.  \label{fig:offset}}

\centerline{\includegraphics[angle=-90,width=1.4\columnwidth, keepaspectratio, clip=true, trim=1.5cm 2cm 0.5cm 5cm]{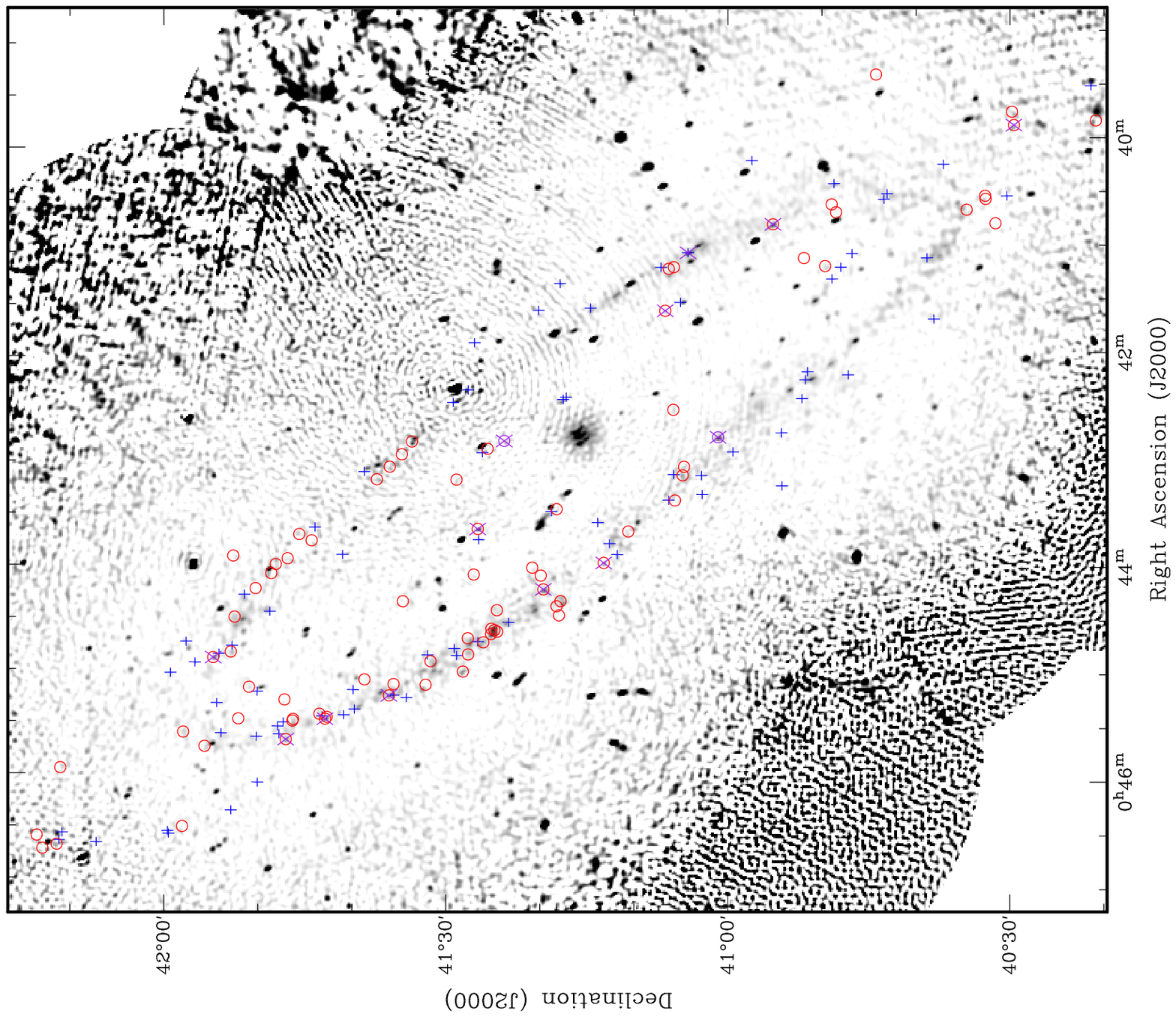}}
\figurecaption{5.}{A radio-continuum image of \M\ from Galvin et al. (2012) overlaid with the positions of SNRs identified by Lee et al. (2014). The red circle represents an all previously identified SNRs that were presented in Lee et al. (2014), while the blue cross represents new SNR candidates identified by Lee et al. (2014). The purple stars highlight sources common both to our study and those presented by Lee et al. (2014).\label{fig:M31OSNR}}

\centerline{\includegraphics[angle=-90,width=1.4\columnwidth, keepaspectratio, clip=true, trim=1.5cm 2cm 0.5cm 5cm]{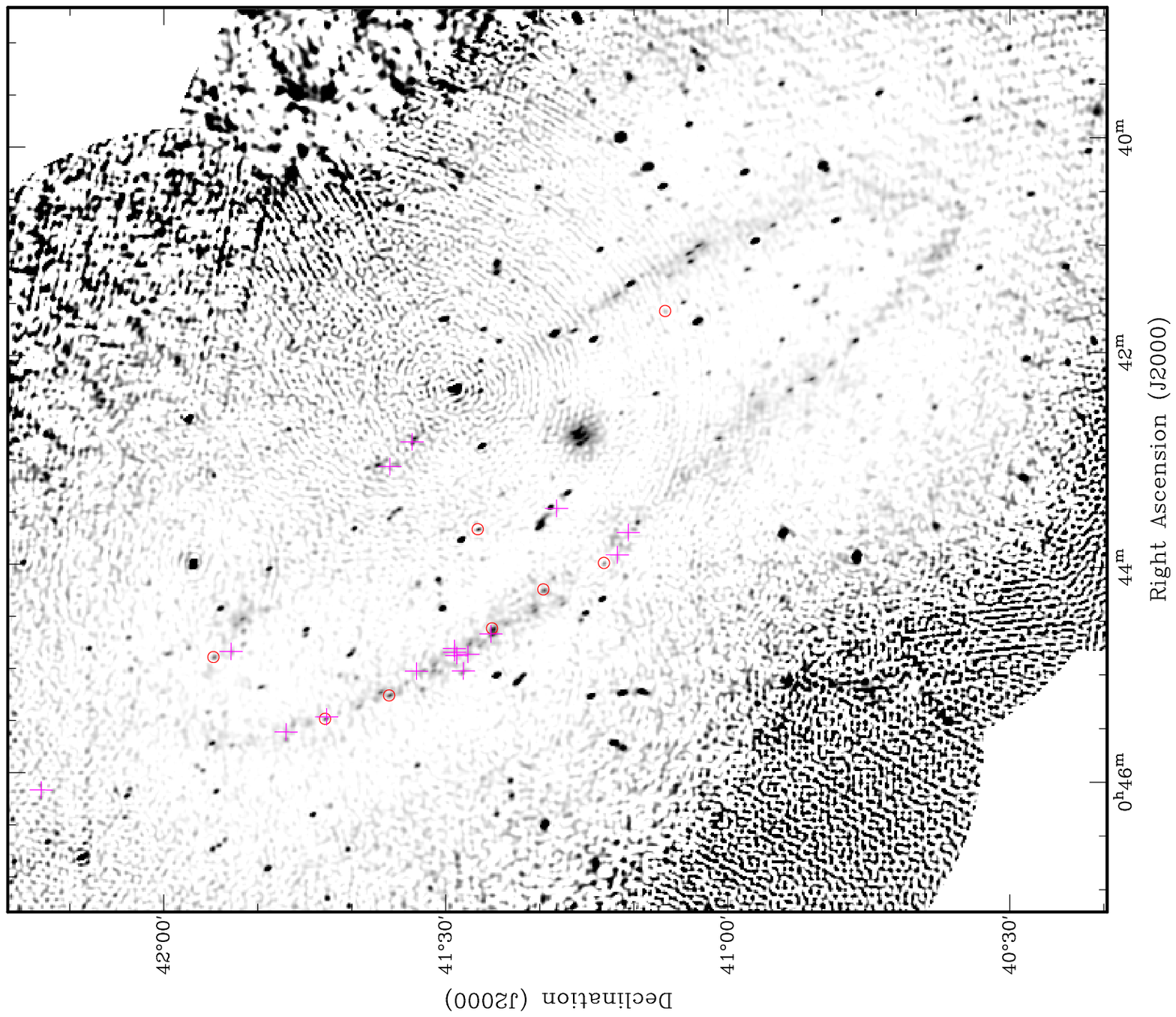}}
\figurecaption{6.}{A radio-continuum image of \M\ from Galvin et al. (2012) overlaid with the positions of SNRs identified by by Braun and Walterbos (1993). The red circle represents sources common to both our study and those listed in Braun and Walterbos (1993), while the purple crosses represent sources in Braun and Walterbos (1993) which did not have a corresponding source in our study. }

\centerline{\includegraphics[width=0.85\columnwidth, keepaspectratio, clip=true, trim=0cm 0cm 0cm 0.5cm]{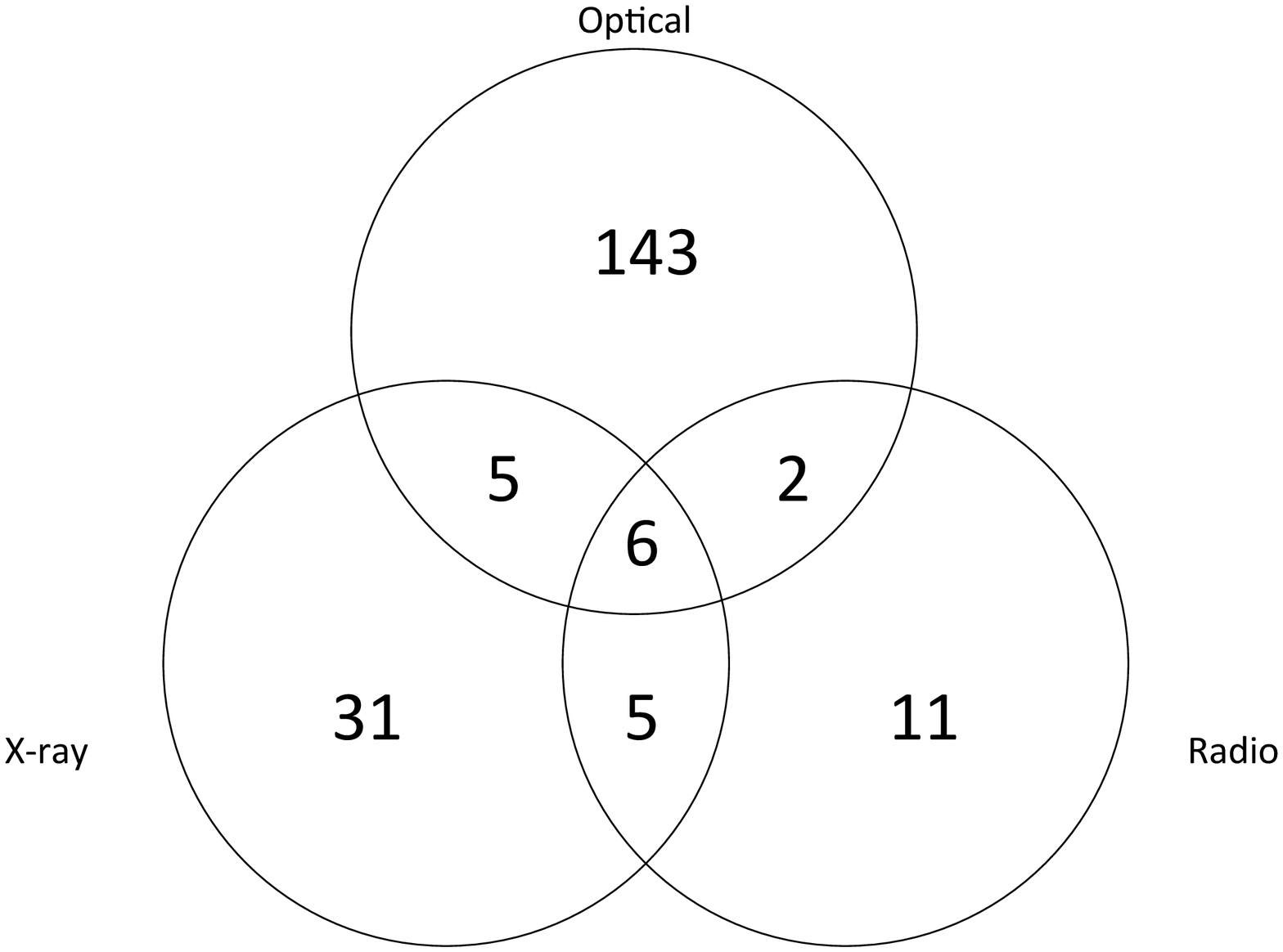}}
\figurecaption{7.}{Venn Diagram showing the intersection of sources from optical (Lee at al. (2014)), radio (Braun and Walterbos (1993)) and X-ray (Sasaki et al. (2012)) catalogues from SNR and SNR candidates sources in the field of \M. \label{fig:venndiagram}}

\begin{landscape}

\tiny{
\begin{longtable}{llllllllllll}
\caption{A summary of all flux density measurements from sources detected in imaged projects. Column~1 is the name of the source, Column~2 and Column~3 are the RA and Dec positions (J2000) of each detected source, and Column~4 is the project in which the source was found. Sources common to multiple project have been assigned a Group ID value, listed in Column~5. The size of each group is listed in Column~6. For each group a mean flux and associated error was calculated and listed in Columns~7 and 8. Columns 9 and 10 are from Gelfand et al. (2004). Column~11 is the spectral index of either Column~4 or Column~8 and Column~10. A source name marked with * indicates that it has an optical counterpart marked as an SNR candidate by Lee et al. (2014).}
\label{table:sources}
\tabularnewline

\hline\hline
Name 	&  RA (J2000)	&  Dec (J2000)			&  Flux 	&  Project &  GroupID 	&  GroupSize 	&  Average 	&  Error 	&  GFlux 	&  GError Flux 	&  Alpha 	\\
		& h m s		& \degr\ \arcmin\ \arcsec\	& (mJy)	&		&			&			& (mJy)		& (\%)	& (mJy)	& (mJy)		&		\\
\hline
\endfirsthead

\hline\hline
Name 	&  RA (J2000)	&  Dec (J2000)			&  Flux 	&  Project &  GroupID 	&  GroupSize 	&  Average 	&  Error 	&  GFlux 	&  GError Flux 	&  Alpha 	\\
		& h m s		& \degr\ \arcmin\ \arcsec\	& (mJy)	&		&			&			& (mJy)		& (\%)	& (mJy)	& (mJy)		&		\\
\hline
\endhead

\hline
\multicolumn{12}{l}{Continued on the next page...} \\
\endfoot
\hline
\endlastfoot
J003645+404524 & 00:36:45.82 & +40:45:24.40 & 7.08 & AC0308-3 &  &  &  &  &  &  &  \\
J003651+404452 & 00:36:51.36 & +40:44:52.40 & 4.81 & AC0308-3 &  &  &  &  &  &  &  \\
J003724+403821 & 00:37:24.96 & +40:38:21.30 & 2.61 & AC0308-3 &  &  &  &  &  &  &  \\
J003730+401239 & 00:37:30.21 & +40:12:39.10 & 4.95 & AC0308-3 &  &  &  &  &  &  &  \\
J003745+402513 & 00:37:45.61 & +40:25:13.50 & 22.56 & AC0308-3 &  &  &  &  &  &  &  \\
J003807+405252 & 00:38:07.88 & +40:52:52.90 & 3.04 & AC0308-3 &  &  &  &  & 11.16 & 0.67 & -0.85 \\
J003813+403100 & 00:38:13.13 & +40:31:00.80 & 5.89 & AC0308-3 &  &  &  &  &  &  &  \\
J003814+403425 & 00:38:14.44 & +40:34:25.70 & 6.53 & AC0308-3 &  &  &  &  &  &  &  \\
\end{longtable}
}
\end{landscape}

\begin{multicols}{2}
{
\acknowledgements{We used the {\sc karma} and {\sc miriad} software packages developed by the ATNF. The National Radio Astronomy Observatory is a facility of the National Science Foundation operated under cooperative agreement by Associated Universities, Inc.}


\references



{Blair}, W.~P., {Kirshner}, R.~P., {Chevalier}, R.~A.: 1981, \journal{Astrophys. J.}, \vol{247}, 879.

Braun, R.: 1990a, \journal{Astrophys. J. Suppl. Ser.}, \vol{72}, 755.

Braun, R.: 1990b, \journal{Astrophys. J. Suppl. Ser.}, \vol{72}, 761.

Braun, R. and Walterbos, R.: 1993, \journal{Astron. Astrophys. Suppl.}, \vol{98}, 327.

Dickel, J. R., Dodorico, S., Felli, M., Dopita, M.: 1982, \journal{Aust. J. of Phys.}, \vol{252}, 582.

{Dodorico}, S., {Dopita}, M.~A., {Benvenuti}, P.: 1980, \journal{Astron. Astrophys. Suppl.}, \vol{40}, 67.

Filipovi{\'c}, M. D., Payne, J. L., Reid, W., Danforth, C. W., Staveley-Smith, L., Jones, P. A., White, G. L.: 2005, \journal{Mon. Not. R. Astron. Soc.}, \vol{364}, 217.

Filipovi{\'c}, M.~D., Haberl, F., Winkler, P. F., Pietsch, W., Payne, J. L., Crawford, E. J., de Horta, A. Y., Stootman, F. H., Reaser, B. E.: 2008, \journal{Astron. and Astrophys.}, \vol{485}, 63.

Galvin, T. J., Filipovi\'c, M. D., Crawford, E. J., Tothill, N. F. H., Wong, G. F., De Horta, A. Y.: 2012, \journal{Serb. Astron. J.}, \vol{184}, 41.

Galvin, T. J., Filipovi\'c, M. D., Tothill, N. F. H., Crawford, E. J., {O'Brien}, A.~N., Seymour, N., Pannuti, T. G., Kosakowski, A, R., Sharma, B.: 2014, \journal{Astrophys. Space Sci.}, in press.

O'Brien, Andrew N.; Seymour, Nicholas; Pannuti, Thomas G.; Kosakowski, Alekzander R.; Sharma, Biswas

Gelfand, J. D., Lazio, T. J. W., Gaensler, B. M., 2004, \journal{Astrophys. J. Suppl. Ser.}, \vol{155}, 89.

Gooch, R.: 1996, in "Astronomical Society of the Pacific Conference Series, Vol. 101, Astronomical Data Analysis Software and Systems V", G. H. Jacoby \& J. Barnes, ed., 80.

{Haberl}, F., {Sturm}, R., {Ballet}, J., {Bomans}, D.~J., {Buckley}, D.~A.~H.,  {Coe}, M.~J.,  {Corbet}, R., {Ehle}, M., {Filipovic}, M. D., {Gilfanov}, M., Hatzidimitriou, D., La Palombara, N., Mereghetti, S., Pietsch, W., Snowden, S., Tiengo, A.: 2012, \journal{Astron. and Astrophys.}, \vol{545}, A128.

Karachentsev, I. D., Karachentseva, V. E., Huchtmeier, W. K., Makarov, D. I.:  2004, \journal{Astron. J.}, \vol{127}, 2031.

Lee, J. H., Lee, M. G.: 2014, \journal{Astron. J.}, \vol{786}, 130.

{Magnier}, E.~A., {Prins}, S., {van Paradijs}, J.,	{Lewin}, W.~H.~G., {Supper}, R., {Hasinger}, G., {Pietsch}, W., {Truemper}, J.: 1995, \journal{Astron. Astrophys. Suppl.}, \vol{114}, 215.

Massey, P., Olsen, K. A. G., Hodge, P. W., Strong, S. B., Jacoby, G. H., Schlingman, W., Smith, R. C.: 2006, \journal{Astron. J.}, \vol{131}, 2478.

{Millar}, W.~C., {White}, G.~L., {Filipovi{\'c}}, M.~D., {Payne}, J.~L., {Crawford}, E.~J., {Pannuti}, T.~G., {Staggs}, W.~D.: 2011, \journal{Astrophys. Space Sci.}, \vol{332}, 221.

{Millar}, W.~C., {White}, G.~L., {Filipovi{\'c}}, M.~D.: 2012, \journal{Serb. Astron. J.}, \vol{184}, 19.

{O'Brien}, A.~N., {Filipovi{\'c}}, M.~D., {Crawford}, E.~J., {Tothill}, N.~F.~H., {Collier}, J.~D., {De Horta}, A.~Y., {Wong}, G.~F., {Dra{\v s}kovi{\'c}}, D., {Payne}, J.~L., {Pannuti}, T.~G., {Napier}, J.~P., {Griffith}, S.~A., {Staggs}, W.~D., {Kotu{\v s}}, S.; 2013, \journal{Astrophys. Space Sci.}, \vol{347}, 159.

{Pannuti}, T.~G., Duric, N., {Lacey}, C.~K., Goss, W. M., Hoopes, C. G., Walterbos, R. A. M., Magnor, M. A.: 2000, \journal{Astrophys. J.}, \vol{544}, 780.

{Pannuti}, T.~G., {Schlegel}, E.~M., {Filipovi{\'c}}, M.~D., {Payne}, J.~L., {Petre}, R., {Harrus}, I.~M., {Staggs}, W.~D., {Lacey}, C.~K.: 2011, \journal{Astron. J.}, \vol{142}, 20.

Payne, J. L., Filipovi{\'c}, M. D., Pannuti, T. G., Jones, P. A., Duric N., White, G. L., Carpano, S.: 2004a, \journal{Astron. Astrophys.}, \vol{425}, 443.

{Payne}, J.~L., {Filipovi{\'c}}, M.~D., {Reid}, W., {Jones}, P.~A., {Staveley-Smith}, L., {White}, G.~L.: 2004, \journal{Mon. Not. R. Astron. Soc.}, \vol{355}, 44.

{Payne}, J.~L., {White}, G.~L., {Filipovi{\'c}}, M.~D., {Pannuti}, T.~G.: 2007, \journal{Mon. Not. R. Astron. Soc.}, \vol{376}, 1793.

{Payne}, J.~L., {White}, G.~L., {Filipovi{\'c}}, M.~D.: 2008, \journal{Mon. Not. R. Astron. Soc.}, \vol{383}, 1175.

Sasaki, M., Pietsch, W., Haberl, F., Hatzidimitriou, D., Stiele, H., Williams, B., Kong, A., Kolb, U.: 2012, \journal{Astron. and Astrophys.}, \vol{544}, A144.

Sault, R. J., Teuben, P. J., Wright, M. C. H.: 1995, in "Astronomical Society of the Pacific Conference Series, Vol. 77, Astronomical Data Analysis Software and Systems IV", R. A. Shaw, H. E. Payne, \& J. J. E. Hayes, ed., 433.

\endreferences

}
\end{multicols}

\vfill\eject

{\ }




\naslov{20~cm VLA RADIO KONTINUM STUDIJA M\,31 -- SLIKE I KATALOG TAQKASTIH IZVORA DR2: OSTATCI SUPERNOVIH}


\authors{T. J. Galvin, M. D.~Filipovi\'c}

\vskip 3mm


\address{University of Western Sydney, Locked Bag 1797, Penrith South DC, NSW 1797, Australia}
\Email{136525304}{student.uws.edu.au, m.filipovic@uws.edu.au}

\vskip.7cm


\centerline{UDK \udc}



\vskip.7cm

\begin{multicols}{2}
{


{\rrm 
Predstav{\lj}amo verziju 2 na{\ss}eg kataloga taqkastih radio-izvora u galaksiji M\,31 na \textrm{$\lambda$=20~cm ($\nu$=1.4~GHz)}. Ukupno smo identifikovali 916 diskretnih radio-objekata pozicioniranih u po{\lj}u M\,31. Upore{\dj}iva{\nj}em na{\ss}eg kataloga sa katalogom \textrm{Gelfand et al. (2004)} na \textrm{$\lambda$=92~cm,} na{\ss}li smo 98 zajedniqkih objekata za koje smo izraqunali spektralne indekse. Ve{\cc}ina (73\%) ovih objekata ima spektralne indekse od \textrm{$\alpha <$--0.6,} iz qega sledi da je netermalna emisija dominantna --- tipiqno za pozadinske galaksije i ostatke supernovih. Tako{\dj}e, istra{\zz}ivali smo i postaja{\nj}e radio-detekcija za 156 poznatih ostataka supernovih detektovanih optiqkim posmatra{\nj}imai i na{\ss}li smo 13 takvih objekata. Upore{\dj}ivanja sa dodatnim optiqkim, radio i \textrm{X}-zraqnim katalozima ukazuju na malu populaciju ostataka supernovih koji se mogu detektovati na svim ovim frekvencijama. 
}

}
\end{multicols}

\end{document}